\documentstyle[12pt]{article}

\begin{document}
\begin{titlepage}
\begin{center}

August 15, 2001     \hfill    LBNL-48665 \\

\vskip .5in

{\large \bf The 18-Fold Way}
\footnote{This work is supported in part by the Director, Office of Science, 
Office of High Energy and Nuclear Physics, Division of High Energy Physics, 
of the U.S. Department of Energy under Contract DE-AC03-76SF00098}

\vskip .50in
Henry P. Stapp\\
{\em Lawrence Berkeley National Laboratory\\
      University of California\\
    Berkeley, California 94720}
\end{center}

\vskip .5in

\begin{abstract}
At least 18 nontrivial correct choices must be made to arrive
at a ``right understanding'' of the world according to quantum
theory.

\end{abstract}

\end{titlepage}

\newpage
\renewcommand{\thepage}{\arabic{page}}
\setcounter{page}{1}
\noindent  {\bf 1. Introduction.}

I have sometimes wondered why it took me forty years to arrive
at what now seems to me to be the right way to understand
the world according to quantum mechanics. Already in 1959 I had
read von Neumann's book, had invented on my own the many-minds
interpretation, and found its fatal flaw, and had written the
first version of ``Mind, Matter, and Quantum Mechanics.''
The answer to my ponderings became evident when I read Ulrich 
Mohrhoff's article [1] ``The world according to quantum mechanics
(or the 18 errors of Henry P. Stapp).'' `Right understanding'
requires at least 18 nontrivial right answers: the probability 
of getting things right by chance is less than one in a quarter 
million. It may be useful to look at these eighteen choices
individually. To this end I shall consider in turn each of the eighteen 
claims made by Mohrhoff, and explain why I chose in each case the option 
opposite to the one leading to Mohrhoff's way of removing efficacious mind 
from our understanding of Nature.

\noindent {\bf 2. Mohrhoff's eighteen claims}

1: ``an algorithm for assigning a probability to a possible outcome
of a possible measurement cannot also represent a state of affairs.''

But a successful probability algorithm must have a basis in a state of affairs.
A theory that provides an explanation of that basis, or at least a rationally
coherent possible explanation of that basis, is more satisfactory than
just the algorithm alone, because an explanation can point beyond what is 
currently known. That is why many scientists seek not just algorithms
but also explanations and understanding.

2.1: ``The introduction of consciousness... [is] gratuitous.''

The empirical basis of science consists of relationships between 
our conscious experiences. Thus an essential part of any physical
theory is a description of the connection between the mathematical 
formulas and the conscious experiences that are both their empirical 
basis and the link to possible applications. For quantum theory, 
as formulated by its founders, the most profound departure from prior  
science was the introduction of `the observer' into the theory in a 
nontrivial way. The founders certainly would have avoided this radical 
innovation if they had been able to conceive of a satisfactory way to do so. 
But they could not. The introduction into the theory of the experiences 
of the observer seemed to be needed to define the `facts' that  
science had to deal with.

2.2: ``Interpretations that grant quantum states the ontological
significance of states of affairs do not satisfy this fundamental
requirement'' (of being consistent with the probabilistic significance
of quantum states.)

The theories of von Neumann, Bohm, and Ghirardi-Rimini-Weber-Pearle 
all grant to quantum states the ontological significance of states
of affairs, or at least aspects of states of affairs, and all are 
constructed to be consistent also with the probabilistic significance 
of the quantum states.

2.3: ``The word that ought to replace `measurement' in any ontological
interpretation of quantum theory is `(property-indicating) fact'. '' 
``dictionar[ies] define `fact' as a thing that is {\it known}
to have occurred, to exist, or to be true; a datum of 
{\it experience},... Should we conclude from this that the editors
of dictionaries are idealists wanting to convince us that the existence
of facts presuppose knowledge or experience. Obviously not. The correct
conclusion is that `fact', ... is so fundamental a concept that it simply
{\it cannot be defined}---the factuality of events or states of affairs
cannot be accounted for,'' 

Dictionaries are usually cited as giving the meaning of a word,
not as evidence that the concept cannot be defined.  The fact that
both Bohr and the dictionary editors turned  to experience and knowledge
in order to characterize `facts' suggests that facts might indeed have some 
sort of experiential underpinning. But that conclusion does not entail 
commitment to idealistic philosophy. It might signal, rather, the failure 
of both idealistic and materialistic philosophy to adequately come to grips
with the nature of `facts'. That those two extreme positions are both 
inadequate is not an unreasonable possibility: philosophers have been 
debating the merits and deficiencies of those two extreme positions since 
the beginning of philosophy. Thus it could be deeply significant that 
the founders of quantum theory found it necessary to define the `facts' 
in a way that was essentially intermediate between the extremes of 
purely matter-based philosophy and purely mind-based philosphy.
Similarly,  von Neumann's attempt to bring rational order to quantum theory,
without renouncing the goal of understanding nature, is built squarely on 
`facts' that are precisely the {\it psycho-physical events} specified 
by Process I. Each such fact has both an experiential aspect, which is 
the very  same `experience or knowledge of the observer' upon which the 
founders of quantum theory built their version, and also a physical aspect, 
represented by the reduction of the quantum state of the brain (and hence 
also the universe) to a form compatible with the experiential aspect of 
that event. This opens the way to a conception of nature that is built 
directly on the mathematics of contemporary physics, is philosophically
intermediate to idealism and materialism, yet is very different from 
classical dualism, in which the two components are not dynamically bound 
together by the basic mathematical structure of the scientific theory 
itself.

2.4:``Any attempt to explain the emergence of facts... must therefore 
be a wholly gratuituous endeavor.''

The ``therefore'' in this claim rests upon acceptance of the assertion
that ` ``fact'' is so fundamental a term that it cannot be defined.'
But if a fact is what is created by a quantum event that contains
both experiential and physical aspects, and certain proposed mathematical 
rules governing the emergence of these facts explains in a concise and
parsimonious way all the successes of classical and quantum physics, then 
that putative explanation of the emergence of facts would not appear to 
be wholly gratuitous.

2.5: ``If the answer (to the question does physics presuppose
conscious observers) is negative for classical physics then it is  
equally negative for quantum  physics.''
 
Our conscious experiences play no dynamical role in classical physics: 
they enter only as passive witnesses to events that are determined by
local physical laws. Whether the same is true in the quantum universe 
cannot be inferred from analogy to the classical approximation. For 
there is no analog in the classical approximation to the lack of 
determination by contemporary laws of quantum theory of which question 
will be posed, or put to nature. That is, there is in quantum dynamics 
not only the indeterminateness associated with the familiar stochastic 
element, but also the need to define, in connection with each actual fact, 
a particular question that was put to nature. This question is not fixed 
by the known laws of quantum theory. So this issue of the ontological 
character of the `facts', and of the process that fixes them, makes 
quantum physics potentially ontologically different from classical physics: 
ideas suggested by the classical approximation need not suffice 
in the real world.

2.6: ``The parameter t on which  the probability depends
is the specified time of this actually or counterfactually
performed position measurement. It refers to the time of a 
position-indicating state  of affairs, the existence of which
is {\it assumed}.''

This  ``assumption'' is the exactly whole problem. It presupposes,
the existence of an essentially classical fact-defining world. That 
assumption is exactly what troubles many physicists who seek a  
rationally coherent theory. How does one reconcile the {\it assumed} 
existence of a classical world of devices with the quantum character
of the atomic constituents of the devices? Bohr stressed that 
``in every account of physical experience one must describe both 
experimental conditions and observations by the same means of communication 
as the one used in classical physics.[2]''  What Bohr and the other 
founders of quantum theory realized was that the fact that scientists
use classical language to describe to their colleagues  what they have 
`done' and what they have `learned' does not necessarily entail the
existence of an essentially classical-type world of localizable facts
that exist and are well defined without any explicit involvement of the
experiential aspect of  nature.  The classical character of the language
we use the describe the world in which we live could arise
from its utility and survival value, rather than from an actually
existing world of well defined localized classical-type facts 
detemined by purely local physical laws that never need acknowledge 
the existence of experience.

2.7  ``$p(R,t)$ is not associated with the possibility that all of a sudden,
at time $t$, the particle `materializes' inside $R$. It is the probability
with which the particle is found in $R$ {\it given} that at time $t$
it is found in one of a set of mutually disjoint regions... [that includes
R].''

Yes. But what does ``found'' mean? What is the condition for a 'fact' to
exist at time t when it did not exist slightly earlier. In real life, as 
in science, what is `found' depends on what is looked for, and how it is 
looked for. Quantum theory is naturally concordant with that idea. 
`Looking for' becomes a fundamental process that is the proper subject 
of scientific theory.  Why must we stick to extratheoretic facts of an
essentially  classical type? Must the answer to the basic question 
of the nature and origin facts rest on ideas drawn classical physics, 
a theory known to be false, or might it come rather from a rationally 
coherent conception of nature built directly on the mathematics that 
correctly and seamlessly accounts for ALL the known empirical data, 
including those explained by classical physical theory. Must it necessarily 
remain forever true that ``Before the mystery of...the existence of facts
...we are left with nothing but shear dumbfoundment.'' Or can the effort 
to bring the mathematical laws of quantum theory into close coordination
with a conception of reality  give us some toe-hold on a path 
toward understanding the emergence of facts?

3.1: ``A possibility is not the kind of thing that persists
and changes in  time.''

But a propensity, or disposition [my words] can be a function
of time. And the propensities for various events to occur can change
when some event occurs: the propensity for an earthquake to occur
is higher when an earthquake has very recently occurred. In a universe
(or theory) with stochastic elements the notion of time-dependent
propensities can make sense.

3.2 the ``third error... is a category mistake. It consists
of ... treating possibilities as if they possessed an actuality
of their own.'' 

Sir Karl Popper[3] and Werner Heisenberg[4] both recommended treating
quantum probabilities as propensities: i.e., as absolutely existing 
tendencies or `potentia' for quantum (actualization) events to occur.
This notion has a long history in philosophy that dates back to Aristotle.

4: ``the erroneous notion that possibilities are things 
(``propensities'') that exist and evolve in time.''

Within von Neumann's formulation of quantum theory the quantum 
probabilities can be consistently interpreted as propensities that exist 
and evolve in time. 

5: ``astronomical data...support the existence  of  a historically 
preferred family of hypersurfaces, but not of a dynamically preferred
one.''

The empirically preferred surfaces are generally believed to be created 
by `inflation', which is a dynamical process.

6: ``inconsistent combinations of counterfactuals.''

The logic is as follows: If getting a certain  outcome (-)
of a  later measurement L1 entails that the outcome of a certain
earlier measurement R2 is (+), and if getting  outcome (+) of the earlier 
measurement R2 entails that the outcome of a later measurement L2 would 
necessarily be (+), and if the last minute choice later choice between
L1 and L2 cannot influence the outcome of R2 that has already occurred
earlier then one can assert that if R2 is performed earlier then if
L1 is performed later and gives outcome (-), then the outcome (+) would
necessarily have appeared if the last minute choice of the later
measurement had  been to perform L2 instead of L1. 

After contemplation anyone can then see that this claim is 
correct. Logicians have examined it and agree that it is correct. 

7: ``fallacious... proof of... faster-than-light information transfers
of information.''

This claim is based on denying the validity of my reply to claim 6.
That reply stands up under close  scrutiny.  But Mohrhoff's description 
of my proof bears little resemblance to my proof itself: there is no 
butressing: no second proof. There is one concise and rigorous proof [5].

8.1: ``granting free will to experimenters'' does not lead ``to a physical
reality inconsistent with the `block universe'  of SR.

By free will I mean the freedom to will ourselves to act `now' either
in some particular way, or not in that way. The existence of freedom in 
this sense is incompatible with the `block universe' of SR, which says 
that the whole universe is laid out for all time in the way that SR 
(classical special relativity theory) ordains, i.e., such that given the
physical state of the universe for very earlier times there is no possibility 
of choosing now to act in some way or to not act in that way. But the laws 
of quantum mechanics, being dynamically incomplete, do not entail a block 
universe in the way that the deterministic and complete laws of SR do. 
The generally accepted application of the requirements of the theory of
relativity to quantum  theory is to its {\it predictions}: {\bf they} must
conform to the no-faster-than-light requirements of the theory of relativity. 
This is exemplified by Tomonaga-Schwinger relativistic quantum field theory, 
which is completely compatible with instantaneous collapses along spacelike 
surfaces, and in fact {\it demands} such collapse to the extent that one 
accepts the existence of von Neumann's Process I. The laws of 
quantum theory lack the coersive quality of the classical laws that 
entail the block universe.

8.2: ''if the possibility of foreknowledge does not exist,..I can actually
be a free agent.'' 

Not in the sense that I have described, in a universe in which the 
deterministic laws of SR hold. For being free in that sense means being 
free to act in either one of two different ways in a universe with a 
given fixed past. No such freedom exists in a universe that obeys the 
deterministic laws of special relativity. What the person knows, or 
does not know, is not pertinent to that conclusion.

8.3 ``The fact that the future in a sense `already' exists is no reason why
choices made at an earlier time cannot be partially responsible for it.''

The `choices' actually made at an earlier time can certainly be partially 
responsible for what happens later. But in a world that conforms to the 
deterministic laws of classical of SR, and with a fixed early universe, 
that `choice' made at the earlier time is not `free': it could not go 
in either one of two different ways. ``Freedom'',  in this sense is not
compatible with the block universe of SR, but it is compatible with
von Neumann quantum theory, due to the indeterminacy within that  theory
of which question will be put to nature.

9: ``the erroneous notion that the experiential now, and the temporal
distinctions that we  base upon it, have  anything to do with the physical
world.'' 

In the classical limit the `experiential now' indeed has no role. But 
the incompleteness of quantum theory allows the experiential now 
to enter into the causal structure in ways not allowed in
classical limit. Thus arguments based on classical concepts lack
conclusiveness.

The `instantaneous now' plays an essential role in the
relativistic quantum field theory of Tomonaga-Schwinger,
to the extent that one implements the von Neumann Process I, 
which ties the instantaneous now of physics to the experiential
now of psychology.

10: ``There is no such thing as  `an evolving objective
physical world'. ''

There is such a thing in Tomonaga-Schwinger relativistic quantum 
field theory if the von Neumann Process I is implemented in it.

11: ``There is no such thing as an objectively open future and an
objectively closed past.''

There is such a thing in Tomonaga-Schwinger relativistic quantum field 
theory, if the von Neumann process I is implemented in it.

12: The ``attempt to  involve causality at a more fundamental
level'' is an error. ``While classical physics permits the 
anthropomorphic projection of causality into the physical world
with some measure of consistency quantum physics does not.''

I am not projecting the intuition of the causal power of our thoughts 
into the world to explain the deterministic aspect of nature associated 
with the Schroedinger equation. I am concerned rather with explaining
how our thoughts themselves can be causally efficacious. The 
mathematical fact is that the quantum laws allow this, by virtue of the
indeteminacy associated with von Neumann's Process I, (The freedom of
choice in which question to pose---i.e., the basis problem) whereas the
classical laws do not provide or allow an analogous causal efficacy.
In classical theory the future is determined by the past by LOCAL
laws, whereas in quantum theory a present choice of which question to pose
is not detemined by the local laws of contemporary physics.

13: ``Error 12 depends crucially upon ...[the] erronous view that the factual
basis on which quantum probabilities are to be assigned is determined by 
Nature rather than by us.''

One of my very chief points is that quantum theory is incomplete
because the equations of quantum theory do not specify which question
is put to Nature, i.e., which apparatus is put in place, which  `basis'
is used to  determine a `fact'. That role is given to us, the 
observer/participants, by Copenhagen quantum theory. In von Neumann's
formulation this choice is bound up in the psycho-physical
event specified by Process I. That this freedom of choice of basis,
is given to ``us'', is the key point that I exploit to explain how 
mental effort can influence brain activity. Mohrhoff and I seem to
be basic agreement on this essential point.

14: ``QM [is] inconsistent with a fundamental assumption of field
theory.''      

Mohrhoff's views about quantum field theory are definitely 
idiosyncratic. I identify contemporary QM with relativistic
quantum field theory, and, for the description of brain dynamics,
with quantum electrodynamics, supplemented by an external 
gravitational field to represent the effects of the earth's 
gravity. I take physical theories to be created by scientists,
and to be judged in the end by their predictive and explanatory
powers. 

15: ``freedom to choose is a classical phenomenon.''

Mohrhoff's argument is based on his idea that mental choice and effort
must be coersive, rather than dispositional. But quantum theory
allows dispositional causes. And within classical physical theory
there is no possibility of freedom of the kind that I have
described above. But von Neumann quantum theory does allow
that sort of freedom.

16: It is erroneos to claim that ``The objective brain can (sometimes) 
be described as a decoherent statistical mixture of `classically
described brains' all of which must be regarded as real.''

By the objective brain I mean the quantum state defined by
taking the trace of the state (density matrix) of the universe
over all variable other than those that characterize the brain,
and my claim is merely that interaction with the environment
converts this state to a form that can be roughly described
as a mixture of states each of which approximates a classical 
state of the brain, in the sense that in the position-basis the states 
are approximately localized. This mixture includes ALL of these states, 
not just one of them, or some small subset that are all observationally 
indistinguishable. There is therefore the problem of relating this 
state to observation. This is achieved, within the von Neumann 
formulation, by a psycho-physical event described by Process I. This
approach can be described as taking the mathematics of quantum theory
to give valid information about the structure of reality, and
making our idea of nature, and the facts that define it, conform as 
closely as possible to that mathematical description.

17: ``The metaphor of experimenter as interrogator of Nature [is fitting]
within the Copenhagen framework, which accords special status to
measuring instruments, [but is not fitting] in Stapp's scenario,
which accords special status to neural correlates of mental states.''

In the Copenhagen interpretation `the measuring instruments and the 
participant/observer' stand outside the system that is described by 
the quantum mathematics, and they are probing some property of a 
`measured system', which is part of an imbedding quantum universe. 
In the von Neumann formulation the role of `the measuring instuments 
and the participant/observer' is transferred to the 'abstract ego',
and the measured system whose properties are being probed is the
brain of the observer. 

The need for the outside `observer', which includes the measuring
devices, is to specify a definite question, which is represented
by a projection operator $P$ acting on the state of the 
`measured system', even though nothing about that system itself---
or even the whole of the quantum-described universe---defines 
that particular operator $P$, or  specifies {\it when}
that question is `put to Nature.'

A key technical question now arises: To what extent does the
natural Schroedinger evolution of the brain, abetted by the
Environment-Induced-Decoherence (EID), already decompose the quantum
state of the brain into orthogonal branches that correspond to
distinguishable `facts'? The EID tends to create `coherent states',
which have exponential tails, and are not mutually orthogonal.
In fact, it is impossible in principle for a Schroedinger
evolution to define the projection operators $P$ that are
needed in the von Neumann formulas. That fact is undoubtedly
why von Neumann introduced his Process I as a basic process
not derivable from the Schroedinger Process II. There is no way
to get the needed projection operators out of Process II.
A projection operator defines, and is defined by, a 
{\it subspace}. That means that the definition of a projection
operator P must distinguish every vector in the subspace from
vectors that lie outside that subspace and differ from that vector
by infinitesimal amounts. The Schroedinger evolution of an isolated 
system could define and preserve energy eigenstates, but there is no way that
the Schroedinger evolution of a brain interacting with its
environment could specify subspaces associated with
experientially distinct facts without some extra rule
or process not specified by the Schroedinger Process II
itself. This fact is the technical basis of von Neumann's
theory. It is this need for some outside process
to fix the content of the facts, and their times of coming into
being, that opens the door to the efficacy of experience.

17.1: ``The questions the mind can put to the brain, by choosing where to
fix its attention, are always compatible, for the mind does not have to choose
between incompatible experimental arrangements.''

But it does have to define an operator $P$, or, equivalently, a subspace
of the Hilbert space associated with the brain, and this subspace is not
specified by the prior (quantum) state of the brain or the universe.

17.2 ``If attention is drawn to the highest bidder, the highest bidder
is not a component of a mixture of CDBs (Classically Described Brains),
but one among several neural events or activities competing for attention
in one and the same CDB.''

The various CDBs are overlapping (non-orthogonal), and the function
of Process I is to pick out of this amorphous mass of possible states
some well defined subspace associated with a distinct experience. 
Our ruminations about possible choices can run over various consciously
experienced possibilities, but each such experience is associated
with a selection of a subspace from the amorphous collection of overlapping 
possibilities that constitutes the state of the brain at that moment. 
The Schroedinger Process II cannot by itself unambiguously decompose 
the state of the brain into well defined orthogonal branches corresponding 
to distinct CDBs.

18: ``The theory Stapp ends up formulating is completely different
from the theory that he initially professes to formulate, for in the beginning
consciousness is responsible for state vector reduction, whereas in the end
a new physical law is responsible, a law that in no wise depends on 
consciousness.''

That example is not my final theory. It was put forward as a simple
model to show how one could, within the general von Neumann framework 
that I have developed, produce a dynamically complete theory. It does 
this by postulating a new law that resolves in a certain definite way 
the dynamical freedom that I have used to make consciousness efficacious. 

Mohrhoff suggests that this model violates the quantum laws, but  that 
is not true: the extra postulated law controls which questions are asked, 
and when they are asked, and these are precisely the elements that are not 
controlled by the standard quantum laws, and whose indeterminacy, with
respect to those laws, provides the opening for efficacious minds within 
contemporary basic physical theory

The essential points are, first, that some rules that go beyond the 
Schroedinger equation (as expressed within the framework of relativistic
quantum field theory developed by Tomonaga and Schwinger), {\it must}
be added to the Schroedinger equation to tie the quantum mathematics
to the problem of securing statistical connections between human
conscious experiences, and, second, that these added rules are basically
nonlocal, because the evaluation of the formula Trace PS(t) is a nonlocal
operation. (These extra rules are needed not only in the von Neumann
formulation, but in every formulation, including the many minds/worlds
theories, where this dynamical gap is known as "the basis problem". This
basis problem has {\it not} been solved within any framework resting
solely on the Schroedinger equation alone.)
  
The key issue is whether consciousness itself enters the dynamics.
This devolves to the question: What is the ontological source or 
basis of the coersive quality of the laws of nature?

My answer is that with respect to the local dynamical process 
governed by the Schroedinger equation this question is not worth
pursuing: it is the {\it existence} of that mechanical law that 
matters to us, and inquiry into what gives that law its coersive 
quality is unduly speculative. But the same question is far from
meaningless in connection with the collapse process that is associated 
with a human conscious experience. This is because the experiences
associated with these collapses {\it are known to us}, and are in fact 
the only things really known to us. Thus they are proper elements of 
science, and are, in fact, the basis of science. 

I have thus emphasized that the resetting (of the state of the brain) 
associated with a conscious human experience is mathematically and 
dynamically different from the local mechanical (Schroedinger) process,
and I now suggest that the coersiveness of the laws that govern the 
creation of a stream of consciousness comes in part from 
the experiences that constitute that stream: that the experiential
realities are actually doing something that is not done by the
local mechanical laws. Why else would they exist?  

Thus I have noted that some process beyond the local mechanical
process described by the Schroedinger equation is needed to complete basic
contemporary physical theory, and tie it to the empirical facts,
and I am suggesting that the conscious experiences that emerge in this
process are essential causal elements of a nonlocal evaluative process
that is needed to complete the quantum dynamics.

In summary, local mechanical process alone is logically incapable 
picking the question (choosing the basis) and fixing the timings
of the events in the quantum universe. So there is no rational reason 
to claim that the experiential reality that constitutes a stream of
consciousness is not a causal aspect of the dynamical process that 
prolongs or extends this reality, yet lies beyond what the 
local quantum mechanical process is logically able to do.

\noindent {\bf References} 

1. Ulrich Mohrhoff, ``The World According to Quantum Mechanics
     (Or the 18 Errors of Henry P. Stapp)''\\ 
   (http://xxx.lanl.gov/abs/quant-ph/0105097)

2. Niels Bohr, {\it Atomic Physics and Human Knowledge},
    Wiley, New York, 1958.

3. Karl Popper, ``Quantum Mechanics without `The Observer' '',
in  {\it Quantum Theory and  Reality}, ed. M. Bunge,
Springer-Verlag, Berlin, 1966. 

4. Werner Heisenberg, {\it Physics and Philosophy,}
Harper and Rowe, New York, 1958. 

5. Henry P. Stapp, {\it Bell's Theorem without
   Hidden Variables}, Lawrence Berkeley National Laboratory
Report LBNL-46942,  June 18, 2001.\\
http://www-physics.lbl.gov/\~{}stapp/stappfiles.html 

\end{document}